\documentclass[useAMS,usenatbib]{mn2e}
\usepackage{graphics}
\usepackage{epsfig}
\usepackage{graphicx,color}
\usepackage[fleqn]{amsmath}

\newcommand{\mbh}{M_{\rm bh}}
\def\ltsima{$\; \buildrel < \over \sim \;$}
\def\simlt{\lower.5ex\hbox{\ltsima}}
\def\gtsima{$\; \buildrel > \over \sim \;$}
\def\simgt{\lower.5ex\hbox{\gtsima}}
\def\sgra{Sgr~A$^*$}
\newcommand\medd{\dot{M}_{\rm Edd}}

\newcommand\ledd{{L}_{\rm Edd}}

\def\msun{{\,{\rm M}_\odot}}

\def\del#1{{}}

\title[Feedback and accretion in Sgr~A$^*$ and LLAGN]{The role of feedback in accretion on Low Luminosity AGN: Sgr~A$^*$ case study}

\author[Cuadra et al ]{Jorge Cuadra$^{1}$, Sergei Nayakshin$^2$, Q.\ Daniel  Wang$^3$\\
$^1$Instituto de Astrof\'isica, Facultad de F\'isica,
      Pontificia Universidad Cat\'olica de Chile,
            782-0436 Santiago, Chile \\
{E-mail:~} {\rm jcuadra@astro.puc.cl}\\
$^2$Department of Physics \& Astronomy, University of Leicester, Leicester, LE1 7RH, UK\\
$^3$Department of Astronomy, University of Massachusetts, Amherst, MA 01003, USA
}

\begin{document}

\pagerange{\pageref{firstpage}--\pageref{lastpage}} \pubyear{2015}
\maketitle
\label{firstpage}

\maketitle

\begin{abstract}

We present numerical models of the gas dynamics in the inner parsec of the Galactic centre. We follow the gas from its origin as stellar winds of several observed young massive stars, until it is either captured by the central black hole, or leaves the system. Unlike our previous models, we include an outflow from the inner accretion flow.  Two different kinds of outflows are modelled: (i) an instantaneous-response feedback mode, in which the outflow  rate is directly proportional to the current black hole gas capture rate; and (ii) an outburst mode, which is stronger but lasts for a limited time. The latter situation may be particularly relevant to \sgra, since there is evidence that \sgra\ was much brighter in the recent past.
We find that both types of outflow perturb the gas dynamics near the Bondi radius and the black hole capture rate significantly.  The effects persist longer than the outflow itself.
We also compare the effects of spherically symmetric and collimated outflows, and find that the latter are far less efficient in transferring its energy to the surrounding gas near the capture radius.
Our results imply that accretion feedback is important for non-radiative accretion flows not only within but also outside the capture radius. Steady-state Bondi accretion rate estimates that do not account for feedback outflows over-predict not only the accretion rate onto the black hole but also the capture rate at the Bondi radius itself. Finally, the steady-state assumption under which non-radiative flows have been routinely studied in the literature may have to be abandoned if accretion feedback is bursty in nature. 

\end{abstract}


\section{Introduction}

Over the last twenty or so years, an understanding of gas accretion onto black
holes at rates much smaller than their Eddington accretion rates, defined here
as $\medd = \ledd/(0.1 c^2)$, emerged. These flows are very different from
their high accretion rate counterparts that are thought to be well described
by the standard accretion disc model \citep{Shakura73}.  Standard discs are cold, dense,
massive and geometrically thin. In contrast, the low accretion rate flows are:

\begin{enumerate}

\item non-radiative.  The energy generated by viscous torques locally is not
  radiated away \citep{NY94,Narayan95}. This also implies that the discs are
  geometrically thick (aspect ratio $H/R \sim 1$).

\item two-temperature. The low radiative efficiency of the flows is in part
  due to low densities and in part due to ions and electrons decoupling from
  each other thermally, with electrons being much cooler than ions in the innermost region
  \citep{ShapiroEtal76,Quataert98}.

\item wind-launching. Since the flows are non-radiative, they overheat and
  spew out a large fraction of their mass and energy budget in
  outflows emerging from all radii \citep{Blandford99, Begelman2012}.

\end{enumerate}

These so-called Radiatively Inefficient Accretion Flows (RIAF) have been a
success story when applied to observations of Low Luminosity AGN (LLAGN) and
\sgra, the SMBH in the centre of our Galaxy \citep{Melia01,GenzelEG10}, in
particular. The accretion flow region of a super-massive black hole (SMBH) can
be defined by its accretion radius, $R_{\rm A} = {2 G\mbh/ (c_{\rm s}^2 +
  \sigma^2)}$, where $\mbh$ is the black hole mass, and $c_{\rm s}$ and
$\sigma$ are the hot gas sound speed and the host galaxy 1D velocity
dispersion, respectively. Within $R\simlt R_{\rm A}$, the gas dynamics should be
controlled by the SMBH gravity and the gas angular momentum.  For our Galactic centre, 
$R_{\rm A} \approx 0.04\,{\rm pc} \approx 1" \approx 10^5 R_{\rm Sch}$.

Observations of \sgra\ show that only a small, $\sim 0.1-1$\%, fraction of gas
{\em thought} to be accreted at the accretion radius is finally accreted onto
\sgra, confirming point (iii) above \citep{Marrone07, Wang13}, which 
has also been found in state-of-the-art numerical simulations \citep[e.g.,][]{Yuan15}. 
In this paper we call attention to two
important issues: (1) all attempts to understand LLAGN and \sgra\ accretion in
particular assumed a steady-state picture of the accretion flow, and this may
be too simplistic for connecting the data and theory reliably; and (2) Effects
of the outflows driven by RIAFs may actually be felt beyond the circularised
part of the accretion flow, and even beyond the accretion radius.

In this paper we build 3D numerical models of the dynamics of stellar wind accretion onto
\sgra\ in the Galactic Centre, for the first time including the effects of an
outflow launched from the super-massive black hole.  We find that this RIAF
{\em feedback} effect onto the surrounding gas may reduce gas accretion rates
onto SMBH further yet. The complete solution to the surprising dimness of
\sgra\ may therefore include not only what goes on {\em within} the RIAF but
also how the RIAF affects the surrounding hot gas reservoir.

\subsection{Is the accretion flow on to \sgra\ in steady state?}\label{sec:steady?}

The observations of \sgra\ give us a number of specific reasons to think that
the rate at which gas is deposited into the accretion flow region can vary
drastically on relatively short time scales. First of all, the well known star
formation event in the central parsec of the Galaxy
\citep[e.g.,][]{PaumardEtal06} has probably been triggered by a deposition of
a massive, $M\sim 10^{4 - 5}\msun$ gas cloud in that region \citep[e.g.,][]{NC05}. This implies that
\sgra\ has come a full circle from being a local equivalent of a (short lived)
quasar \citep{GuoMathews12,ZubovasEtal11a,ZN12a} some $\sim 5$ million years
ago to a champion under-luminous AGN now.  Indeed, \cite{Mou14} have recently shown
that a very strong outflow ($\dot M \approx 0.02 \dot M_{\rm Edd}$) from \sgra\ could have produced 
the Fermi bubbles.  

Aside from this rather extreme variability, there are indications that
\sgra\ luminosity varied much more recently. X-ray echoes
of the region close to \sgra\ show that it was as bright as $10^{39}$ erg/s
only a few hundred years ago \citep[e.g.,][]{PontiEtal10}.  While it is not
clear at this moment what triggered this change in X-ray luminosity, it is
likely that a significant change in the accretion rate onto \sgra\ is
required.

\sgra\ is fed by accretion of stellar winds from the young stellar cluster
surrounding it.  Many of these young stars are in the Wolf-Rayet phase and
have mass-loss rates of the order of $10^{-5}\msun\,$yr$^{-1}$.  Altogether,
they provide more than enough material to explain the current accretion on to
\sgra\ \citep[e.g.,][]{Quataert04}.  Simulations by
\cite{Cuadra06,CuadraEtal08} indicate that the capture rate of stellar winds
should experience variations of a factor of a few within a time-scale of a
couple hundred years, which corresponds to the orbital period of the closest
wind-emitting stars around \sgra.  Moreover, a fraction of the stellar wind
material may form dense clumps, whose stochastic motion produces spikes in the
accretion rate on even shorter time-scales of a few years.  It is however
unclear how much of that short-time variability will affect the accretion on
to \sgra\ itself, as the variations will be smoothed out over the viscous
time-scale.

The G2 cloud \citep{Gillessen12, Gillessen13, Eckart13, Phifer13} might correspond to one of the
cold clumps predicted by the simulations of \cite{CuadraEtal08}.  Such a cloud, if tidally
disrupted by the black hole, would likely increase \sgra's accretion rate \citep{Schartmann12}.
Even though the capture rate at the Bondi radius would increase only slightly
\citep{Anninos12}, this would be accretion of denser material with a coherent
angular momentum, so it could  change the accretion mode
and result in a higher accretion rate at the event horizon.  During the next
decade we may witness a significant change on \sgra\ accretion flow that could
produce an outflow and will certainly illuminate our understanding of low
accretion rate flows.

There are alternative models for G2 that identify this source with a star that loses large amounts of mass, perhaps being partially disrupted by \sgra\
\cite[e.g.,][]{MurrayClay12, Scoville13, Ballone13}.  \cite{Guillochon14} in particular
make the case that the G2 cloud formed out of the condensation of debris material from the partial disruption of a star.
  On average, every decade or so, a clump from the debris would reach the inner region.
  Moreover, \cite{Pfuhl15} recently reported the orbit of another gas cloud (G1), which is remarkably similar to that of G2, and that would have preceded it by 13\,yr.
 The mass-loading of the inner accretion flow would then be a relatively frequent feature of \sgra, and not a 
  one-off event.

\section{The numerical model}

We use the numerical model for the gas dynamics in the Galactic centre
developed by \cite{Cuadra05, Cuadra06, CuadraEtal08}, which is based on the
well known {\sc gadget-2} \citep{Springel05} code used for cosmological
simulations.  This is an N-body plus smoothed particle hydrodynamics (SPH, see
e.g., \cite{Springel10}) code in which stellar winds are emitted by the young
massive stars that orbit at several arcsec from \sgra.  

The code follows the gravitational and hydrodynamical interactions of the gas
in the potential of \sgra\ and its stellar cluster.  Adiabatic processes and a
radiative cooling function are included.  
The effect of viscosity is not included, as our simulations do not reach the actual accretion flow.
We have developed two different
models for \sgra's outflow, which we implemented on top of the
\cite{CuadraEtal08} set-up and detail in \S~\ref{sec:out}.

\subsection{Stellar winds and accretion}

In the simulation we include the 30 stars that have been identified as having important mass-loss rates.  The stars follow Keplerian orbits around \sgra, which are 
 directly, although not
completely, constrained by observations  \citep{PaumardEtal06}.  
In \cite{CuadraEtal08} we tested several stellar orbital configurations that were consistent with the
observed stellar 2D positions and 3D velocities, and found that they all yield
qualitatively similar results.  In this study we concentrate on the effect of
the outflow from \sgra, thus we use only the `1disc' stellar configuration.
In this set-up, roughly half of the stars are given the $z$ coordinate (along the line of sight) defined by 
\cite{Beloborodov06}, which puts them in a well-defined disc of $10\deg$ thickness.
The rest of the stars are given the $z$ coordinate that minimises the eccentricity of their orbits, resulting in a more isotropic distribution.  As a result, none of the stars gets closer than $1.5"$ from \sgra.  The simulations are started with the stars along their orbits, but 1100\,yr in the past.  When we present results referring to ``the present epoch", that corresponds to having ran the model for 1100\,yr until the orbits have reached their current observed positions.  That time is long enough for the gas dynamics to have reached a quasi-steady state  \citep{CuadraEtal08}.

The wind properties of each star, namely, their mass-loss rates and wind velocities, are obtained from their spectra, following \cite{Martins07,CuadraEtal08}.  The mass-loss rates are in the range $5\times10^{-6}$--$1\times10^{-4}\,\msun\,$yr$^{-1}$, and the wind velocities are in the range 600--2500 km/s.  Notice that this is a very different regime to that studied for other LLAGN, where the stellar winds are those of AGB and red giant stars, and therefore much slower \citep[e.g.,][]{Shcherbakov14}.  Notice also that these stellar wind velocities are comparable to the stellar orbital velocities around \sgra, so both components are relevant and are properly included in the models.

Accretion on to \sgra\ is measured from the simulations as the rate at which matter enters a 
$0.1"$ sink radius.  By default, that matter is taken out of the computational domain (but see below).
In \cite{CuadraEtal08} we found that the captured gas had a circularisation radius of $\simlt0.05"$, inside 
the inner boundary of the calculations\footnote{See also the recent study by \cite{BuEtal14}, which suggests the material might not circularise at all.}.
Therefore, our simulations are focused on characterising how the onset of the accretion flow may be 
affected by an outflow from the innermost regions, where the bulk of the gravitational energy is released, rather than 
on providing a comprehensive model of the RIAF itself.

\subsection{Outflows from RIAFs}\label{sec:out}

\cite{Blandford99} pointed out physical reasons for the formation of outflows from
RIAFs. Sadowski et al (2013) have recently simulated accretion and outflows
from non-radiative flows, including the effects of strong magnetic fields, and
found that the outflow generally breaks onto a collimated jet and a wide angle
wind. The former carries most of the energy while the latter carries most of
the mass away from the SMBH. 

We lack the resolution to model gas flows on such small scales in this paper,
therefore we concentrate mostly on  isotropic outflows, which may result from 
an event that is not directly related to the hot accretion process (e.g., tidal disruption 
of a star or planet or a cold dense gas cloud, \cite{NayakshinSS12,Moscibrodzka12}).   In
addition, due to variability in \sgra\ feeding, it is not clear how a weak jet
component (including its orientation) varies with time. It is quite possible
that despite being collimated, the jet shocks easily on the surrounding gas in
the vicinity of \sgra\ and hence inflates quasi-spherical hot bubbles. Such an
outcome is found in the simulations of jet interaction with the ambient ISM
of the host galaxy \citep{WagnerEtal12b,WagnerEtal13a}.
Still, we present a few tests in which the outflow is collimated, with either 
variable or fixed orientation.
  
Outflows in a variety of astrophysical objects are launched at velocities of a
few times the escape velocity. For the accretion disc this corresponds to a
few times the local Keplerian velocity, $v_{\rm K} = \sqrt{GM/R} \approx 2\times10^8\, {\rm cm\,s}^{-1}\, (R/0.1")^{-1/2}$. 
Since the bulk of the mechanical energy is likely generared within our sink radius, the expected outflow velocity
ranges from $2,000\,$km/s to $\sim 0.1 c$ for the broad component. The jet component
could of course drive a relativistic outflow. Such high outflow velocities
would pose a serious numerical challenge to us here, so the maximum
outflow velocity we model is $10^4$~km/s. This does not present a serious
limitation to our results because it is primarily the energy outflow rate that
is important for gas dynamics in the problem at hand, as long as the shocked
outflow is non-radiative.  On the  other hand, the energy deposit could happen entirely outside the present simulation box, which is consistent with the lack of evidence for any significant jet-like feature or its interaction with the surrounding.

In the next subsections, we detail the two different models we use to include an outflow from the vicinity of \sgra\ in the simulations.

\subsubsection{Instantaneous feedback}\label{sec:inst}

We first implemented  an instantaneous feedback mode, in which there is a mild outflow which responds to the accretion rate and is present throughout \sgra\ recent history.  This is achieved numerically by expelling, instead of accreting, all particles that reach the sink radius, which in this study is kept fixed at $0.1" \approx 10^4R_{\rm Sch}$.  Note that black holes in RIAFs do accrete some gas, of course, but the accreted gas fraction is thought to be quite small, e.g., $< 0.01-0.1$ of what enters the RIAF at the Bondi radius \cite[e.g.,][]{Blandford99}.  This justifies that in this model we simply expel all of the gas. 

The particles are expelled with a constant velocity that we treat as a free parameter in the range $10^3 - 10^4\,$km$\,$s$^{-1}$.  The mass-loss rate is not a parameter of this model -- it is by definition equal to the instantaneous accretion rate at the sink radius.  In most simulations the outflow is isotropic, but we also present two cases in which the outflow is collimated:  a `fluctuating jet' case in which each particle is expelled in the direction of its angular momentum vector at the moment of capture, and  `bipolar' cases in which particles are expelled in a fixed bipolar cone with half-opening angle of $15\deg$.

\begin{figure*}
\begin{picture}(600,250)
 \put(0,23){ \includegraphics[width=0.52\textwidth]{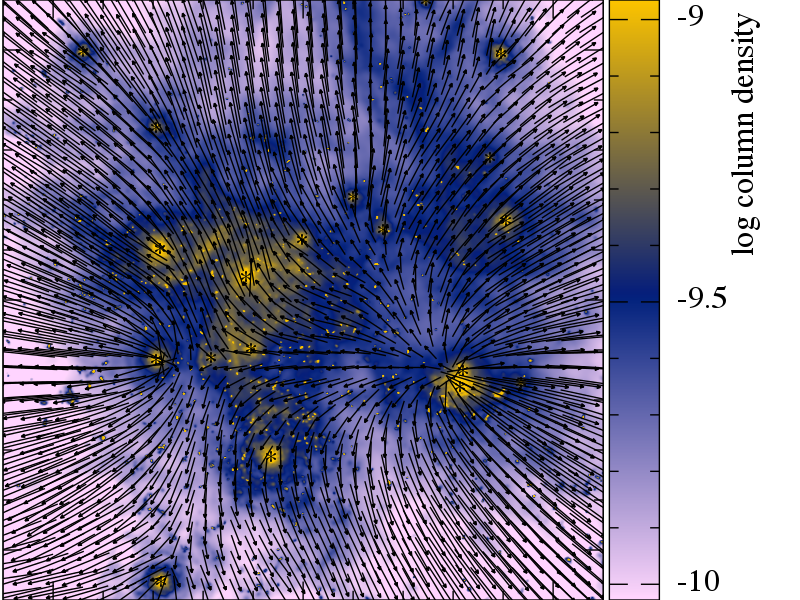}}
 \put(250,0){ \includegraphics[width=0.55\textwidth]{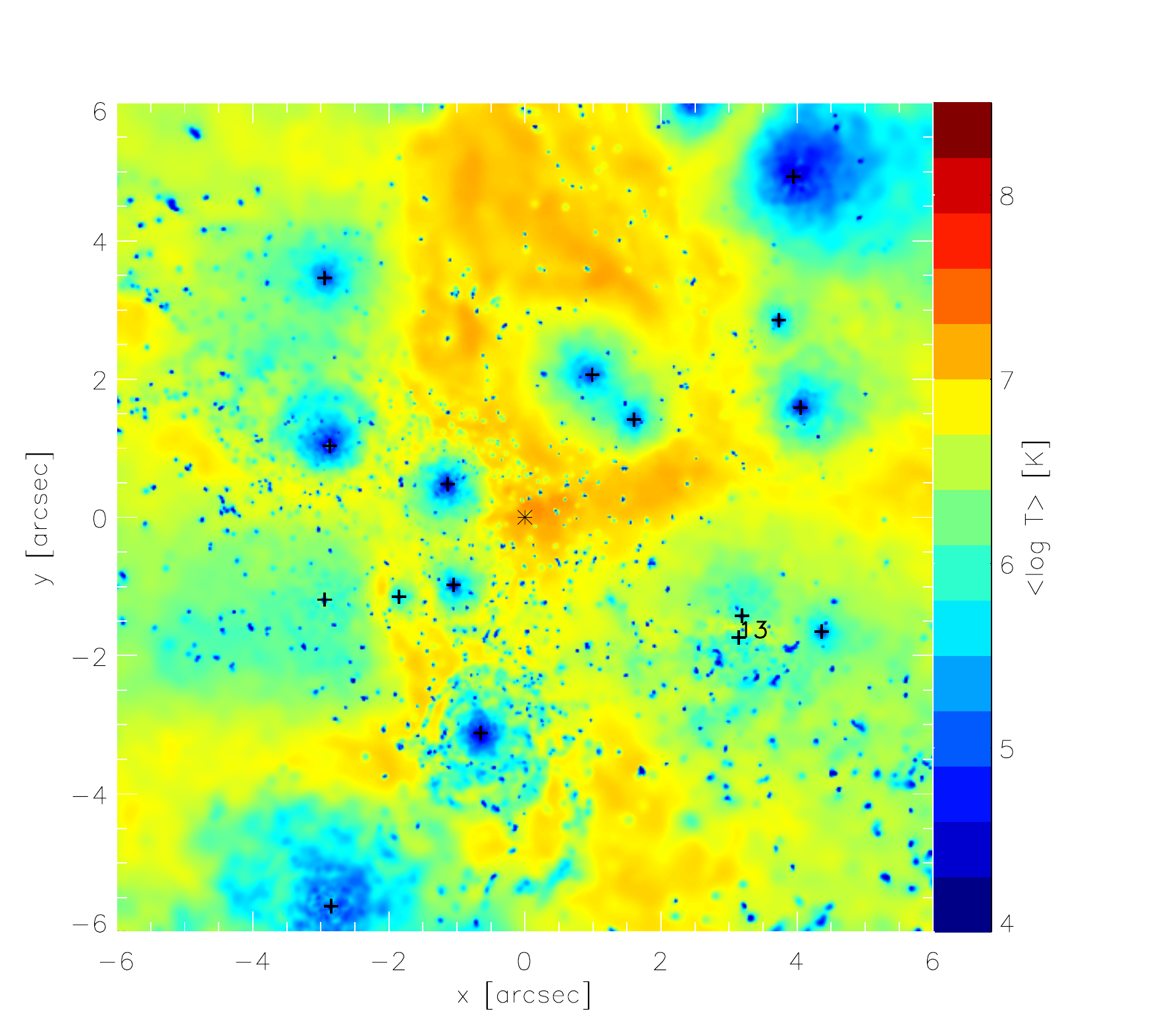}}
\end{picture}
	\caption{Column density and temperature maps of the control simulation.  The maps are $12"$ a side and centred on the black hole.  On top of the left panel, the projected velocity field is shown with vectors.  On the right panel, \sgra\ is shown with an asterisk and stars are shown with small crosses.  The IRS13E group is labelled with the number 13.}
	\label{fig:control_densmap_vel}
	\label{fig:control_tempmap}
\end{figure*}

\subsubsection{Outburst}\label{sec:ob}

We implemented a second mode of outflow in which we assume that for a period of time lasting $300\,$yr and ending $100\,$yr ago \sgra\ was in a state of higher activity, as hinted by X-ray observations \citep{Gando13}.    We will follow the ADIOS model \citep[their example (vi)]{Blandford99}, and set a fiducial run with  outflow velocity $v=5\times10^8\,$cm$\,$s$^{-1}$  at a radius of $R\sim10^4R_{\rm Sch}$. The outflow rate is $\dot M_{\rm out} = 10^{-4}\msun\,$yr$^{-1}$.  These are the values that we would expect for a $\dot M_{\rm in} \sim10^{-5}\msun\,$yr$^{-1}$ accretion rate at the innermost stable orbit, following a $\dot M_{\rm in} \propto R^{-3/4}$ power-law dependence, as proposed by \cite{ Blandford99}.  Moreover, these numbers give a mechanical power output of $\sim10^{39}\,$erg$\,$s$^{-1}$, which coincides with the luminosity inferred for that period.    In the same way as we modelled the stellar winds \citep{Cuadra06}, the outflow has a low temperature of $10^4\,$K, meaning that its energy budget is dominated by its kinetic energy, which gets thermalised once the outflow shocks against the stellar winds.

Numerically, the outflow is produced by temporarily turning the sink particle that represents the central black hole into a source particle, such as the stars.  During this active phase, new gas particles are created around the black hole, at the rate and with the initial velocity quoted above.  In most simulations the outflow is isotropic, but we also present a `bipolar' case, with a fixed direction and a half-opening angle of  $15\deg$.

\section{Simulations and results}

\subsection{Control run}

We start by presenting results from the control run, in which no outflow from the inner region is included  and the gas that reached the sink radius was simply accreted by the black hole.  This run is basically equivalent with the `1disc' model presented by \cite{CuadraEtal08}.

In the left panel of Fig.~\ref{fig:control_densmap_vel} we show a surface density map of the control run at the present epoch (i.e., the stars are at their currently observed positions).  Notice the overall trend of density decreasing with radius.  Besides that, there are local density enhancements around the stars, especially in regions with high stellar densities, where stellar winds collide, and around stars with large mass-loss rates and low outflow velocities, which naturally produce higher gas densities.  On top of the density colour map, the arrows show the projected velocity field.  Notice that, even in this control run, in most of the map the velocity field can be well-described by a roughly radial outflow, with  perturbations due to the effect of some powerful stellar outflows.  For this reason, to analyse the effect of the accretion-produced outflow on the velocity field, in this study we will rely on velocity profiles rather than maps.

In the right panel of Fig.~\ref{fig:control_tempmap} we show a temperature map of the same snapshot of the control run.  To avoid the hot regions to dominate the averages, each point shows the mass-weighted average of the temperature logarithm along the line of sight, i.e., $<\log T> = \int dz \rho \log T / \int dz \rho$, with the integral defined over a range of the same size as the map.  In the figure we can see a large range of temperatures, going from the imposed temperature floor of $10^4\,$K up to $\sim10^8\,$K.  The cold gas corresponds to the stellar winds before they collide (when they shock and thermalise), and to dense clumps that form in some of those collisions, due to thermal instabilities (see \cite{CuadraEtal08} and Calder\'on et al, in prep).  Most of the shocked stellar winds retain a high temperature and are distributed smoothly over the simulation domain.

\subsection{Instantaneous feedback}

Here we present the results of the simulations in which we implemented an instantaneous feedback, i.e., there was an outflow proportional to the inflow rate at the sink radius of the simulations, $0.1"$.  We use four different values for the outflow velocity, namely $ v_8 = v_{\rm out} / 10^8\,$cm$\,$s$^{-1} = 1, 2, 5, 10 $.  

\begin{figure}
	\centering \includegraphics[width=0.5\textwidth]{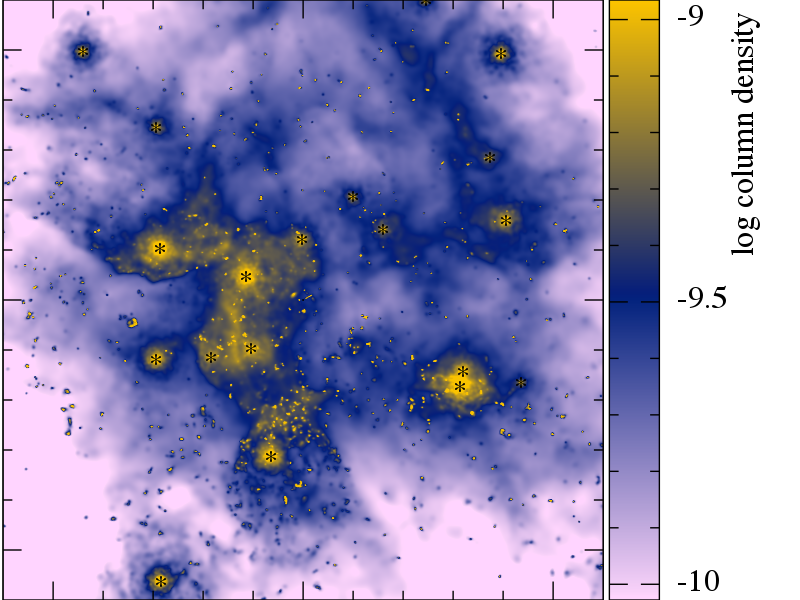}
	\caption{Column density map of the simulation with instantaneous feedback and $v_8=10$.  The map is $12"$ a side and centred on the black hole.  Notice the lower density in the black hole vicinity.}
	\label{fig:if_1e9_densmap}
\end{figure}

We first present in Fig.~\ref{fig:if_1e9_densmap} a density map of the model with the highest outflow velocity, $v_8 = 10$, taken at the current time, so it is directly comparable to the left panel of Fig.~\ref{fig:control_densmap_vel} from the control run.  Notice that the gas morphology is practically the same, except for a small decrement in the inner $1"$ region.  Within the range of velocities we are exploring, the instantaneous feedback outflow mode does not produce any large-scale signature, like cones or bubbles, and its effect is circumscribed to within the Bondi radius.

\begin{figure}
	\centering \includegraphics[width=0.5\textwidth]{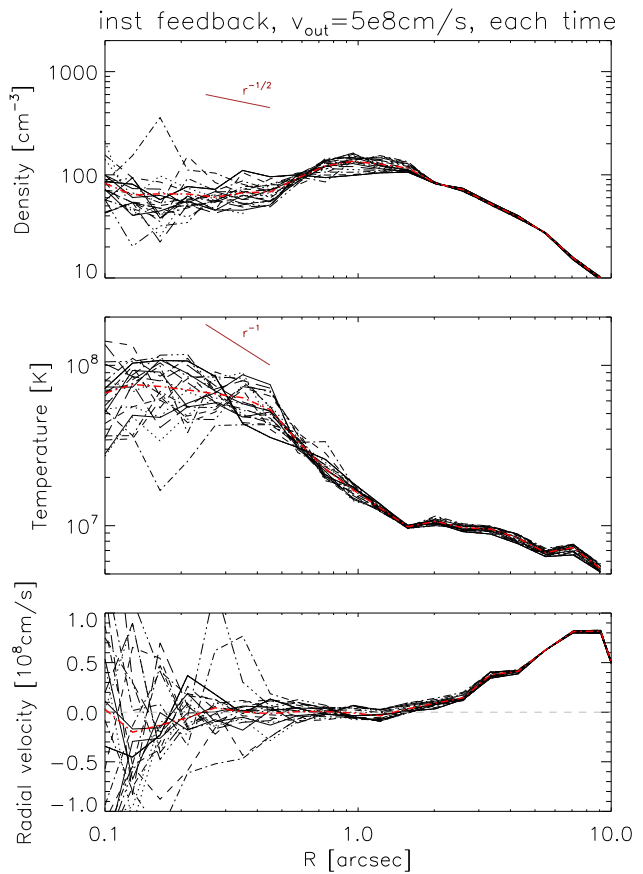}
	\caption{Density (top), temperature (middle), and radial velocity (bottom) profiles of the gas for the  run using the instantaneous feedback mode with $v_8=5$.   Black lines correspond to different snapshots of the simulation, and the red line shows their average. The brown power-law lines in this and the following plots show the observational constrains obtained by \citep{Wang13}, namely, $n\propto r^{-3/2+s}$ with $s\sim1$, and $T\propto r^{-\theta}$ with $\theta \sim 1$.}
		\label{fig:if_5e8_profiles_all}
\end{figure}

Figure~\ref{fig:if_5e8_profiles_all} shows the radial profiles of density\footnote{Defined simply throughout the paper as the mass density $\rho$ divided by the Hydrogen mass.}, temperature and radial velocity of the gas from the model with $v_8=5$.  The different lines correspond to snapshots taken at different times in an interval of $\pm 61\,$yr from the present time, while the red thick line shows their average.  The average represents well the state of the system for $r \simgt 0.5"$, but the conditions of the gas in the very inner region are highly time-variable and can deviate significantly from the average, in particular the radial velocity.  For the rest of the paper, we will use time-averaged profiles in order to compare different simulations, but the reader must be aware that they might deviate somewhat from the state of the system at any given time.

\begin{figure}
	\centering \includegraphics[width=0.5\textwidth]{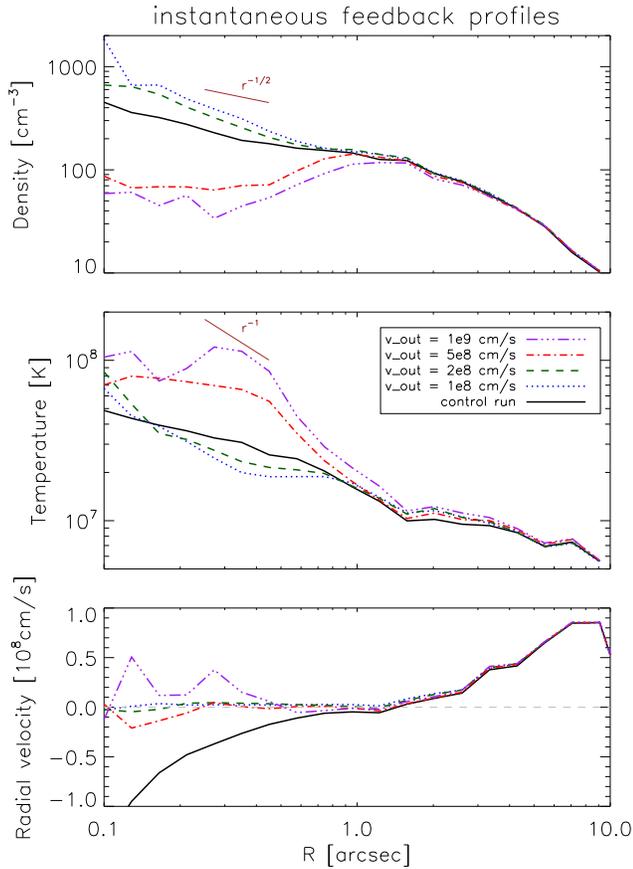}
	\caption{Density (top), temperature (middle), and radial velocity (bottom) profiles for the different runs using the instantaneous feedback mode.  The black solid lines show the profiles for the control run, in which no outflow was introduced; the other lines show the profiles for the runs with outflow -- see inset legend for details.   }
	\label{fig:if_profiles}
\end{figure}

Figure~\ref{fig:if_profiles} shows the time-averaged radial profiles  for the different simulations.  This data visualisation allows to compare among the models better than maps, since the differences are typically only important within the inner $1"$.  
The profiles are built averaging over 21 snapshots that encompass the present epoch, $\pm 61\,$yr.
Notice that the profiles for the control run differ from simple spherical models \citep[e.g.,][]{Quataert04} due to the fact that the accreted material  originates from a finite number of stars, which are moreover preferentially oriented in a plane \citep{Cuadra06}. 
For the cases with $ v_8 = 1, 2$ there is an enhancement in the inner density, as the material that would otherwise be accreted is instead launched outwards at the accretion radius. However, since the launching velocity is lower than the escape velocity from the region, the material then stays in that region and therefore there is a density enhancement around the sink radius. In contrast, in the cases $ v_8 = 5, 10$, the expelled material is able to escape the inner region and the density does decrease compared to the control run.  Overall there's a clear trend of higher inner densities for lower outflow velocities.  The temperature profiles show higher inner temperatures for higher outflow velocities, due to the additional energy which is injected to the gas which gets quickly thermalised.  Notice how the region outside the inner arcsecond is almost completely dominated by the stellar winds -- the outflow from \sgra\ only manages to slightly heat up the gas in that region.  The two upper panels also show the observational constrains derived from the recent {\it Chandra} observations of the region around \sgra, namely,%
\footnote{Notice that the parameters are not determined independently, with $s \sim  \theta$.}
 $n\propto r^{-3/2+s}$ with $s\sim1$, and $T\propto r^{-\theta}$ with $\theta \sim 1$ \citep{Wang13}.  The density profiles favour models with slow or no outflow, while the temperature profiles favour the models with faster outflows.  In conclusion, none of these models seem to be consistent with the current data.

The lower panel of Fig.~\ref{fig:if_profiles} shows the radial velocity profiles.  The region where the stars are, $R\simgt1"$, shows how the stellar winds dominate the dynamics there, and that most of the gas escapes from this region.  In the inner part of the flow, we would expect the radial velocity to be null in steady-state, as the outflow rate from the inner accretion flow is by construction equal to the inflow rate.  However, they do not cancel out exactly due to fluctuations (see Fig.~\ref{fig:if_5e8_profiles_all}).  This effect is stronger when the outflow velocity is higher, but notice that the fluctuations are at the level of a few percent of the outflow velocity only.   This statistical noise also shows up as fluctuations in the density and temperature profiles for the $v_8=10$ run.  An average over a much larger number of snapshots would be needed in order to smooth these curves out, but that is not needed to draw our conclusions.

\begin{figure}
	\centering \includegraphics[width=0.5\textwidth]{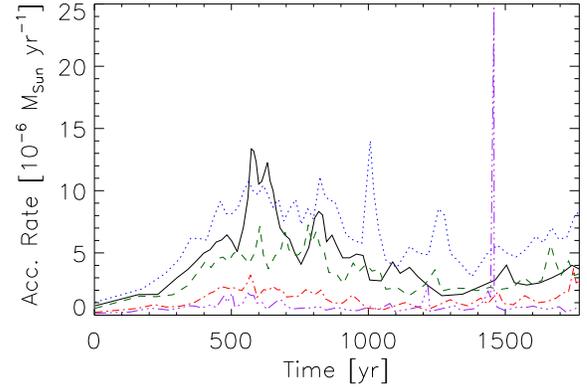}
	\caption{``Capture rates" as a function of time for the different runs using the instantaneous feedback mode.  The line meanings are the same as in Fig.\ref{fig:if_profiles}.}
	\label{fig:if_accretion}
\end{figure}

Figure~\ref{fig:if_accretion} shows the rate at which gas enters the sink radius.  For the control run, this corresponds to the accretion rate, as that gas is actually accreted by the sink particle representing the black hole and no outflow is produced at the inner boundary.  For the instantaneous feedback runs the gas that reaches that radius is expelled, so the ``capture rate" in 
these simulations describes simply the rate of gas entering the $\sim10^4 R_{\rm Sch}$ region rather than \sgra accretion directly, which in our model is, by definition, zero. However, it is expected that in reality a fraction of that captured gas be accreted onto the black hole eventually. Therefore, 
 the plotted capture rates do give an idea of the effect of the outflow on regulating the accretion flow boundary conditions.  
 
 From the figure, it is clear that higher outflow velocities result in smaller capture rates at the sink radius.  The simulation with the lowest velocity, $v_8 = 1$, even shows a higher capture rate than the control run -- as that velocity is lower than the escape velocity, the expelled gas remains in the region and can be re-captured by the black hole.  In contrast, in the control run the captured gas is accreted by the black hole instead of expelled, and therefore can only be ``captured" once.
  It is worth pointing out that all the runs show nearly identical average physical conditions for the gas at $R_{\rm A}$ (see Fig.~\ref{fig:if_profiles}), but they differ for up to an order of magnitude in terms of the capture rate at our sink radius, in a ten times smaller scale.  Clearly the Bondi formula cannot be applied when we know that outflows should be produced from within the accretion radius.  
 
 Notice also that all capture rate curves are variable, even the one from the control run.  This variability is due mostly to the fact that the gas originates from nearby stars, which have non-circular orbits with periods as low as a couple hundred years.  Over such time-scales the stellar configuration changes and that influences the rate of gas that reaches the black hole \citep{CuadraEtal08}.  Moreover, there is formation of cold ($T\simlt10^5\,$K) clumps of gas, which sometimes are accreted, as seen at $t \approx 1400\,$yr for the $v_8 = 10$ run.

\begin{figure}
	\centering \includegraphics[width=0.5\textwidth]{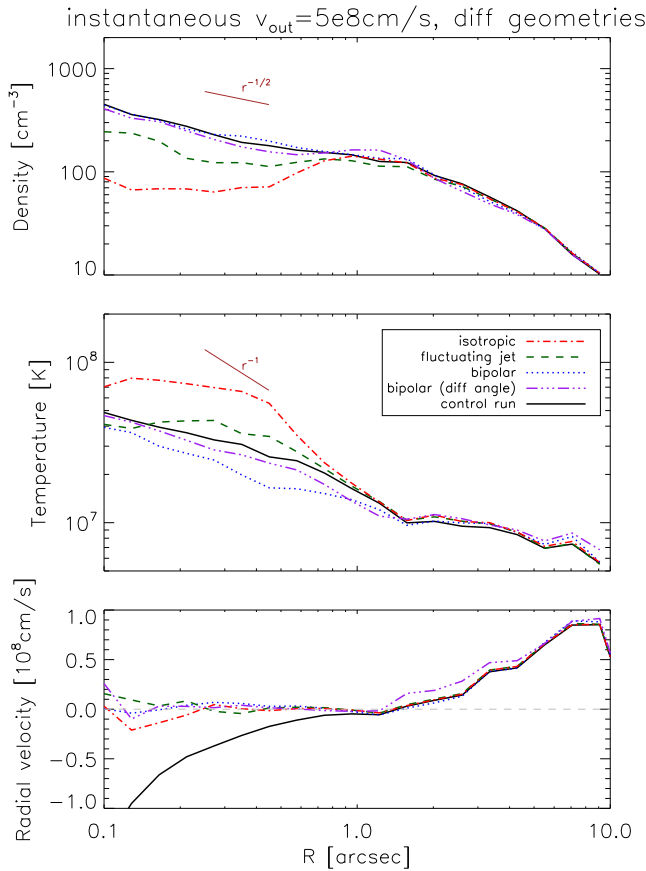}
	\caption{Radial profiles of different gas properties (as in Fig.~\ref{fig:if_profiles}) for the different runs using the instantaneous feedback mode with $v_8=5$ and different outflow geometries.  The black solid lines shows the profiles for the control run, in which no outflow was introduced; the other lines show the profiles for the runs with outflow -- see inset legend for details.}
	\label{fig:if_profiles_bip}
\end{figure}

\begin{figure}
	\centering \includegraphics[width=0.5\textwidth]{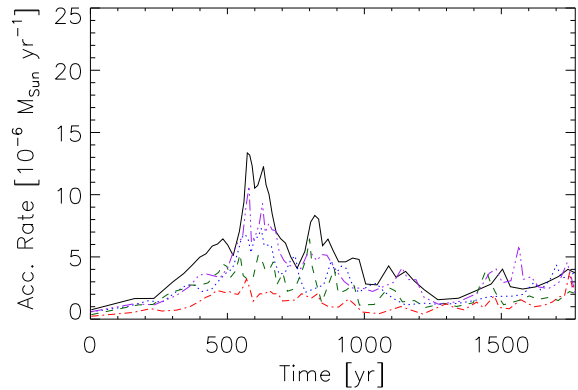}
	\caption{``Capture rates" as a function of time for the different runs using the instantaneous feedback mode with $v_8=5$ and different outflow geometries.  The line meanings are the same as in Fig.\ref{fig:if_profiles_bip}.}
	\label{fig:if_accretion_bip}
\end{figure}

Figures \ref{fig:if_profiles_bip} and \ref{fig:if_accretion_bip} show the results for runs with $v_8=5$ but with different outflow geometries.  The isotropic and control runs are the same presented above, the `bipolar' runs correspond to outflows which are confined to a fixed cone with a half-opening angle of $15\deg$, and the `fluctuating jet' has each particle ejected in the direction of its angular momentum at the moment of crossing the inner boundary.  The results show that the beaming reduces the outflow effect:  the density profile and the capture rate for the `fluctuating jet' and `bipolar' cases take intermediate values between the control run and the isotropic run with the same outflow velocity. 

\subsection{Outburst}

\begin{figure}
	\centering \includegraphics[width=0.5\textwidth]{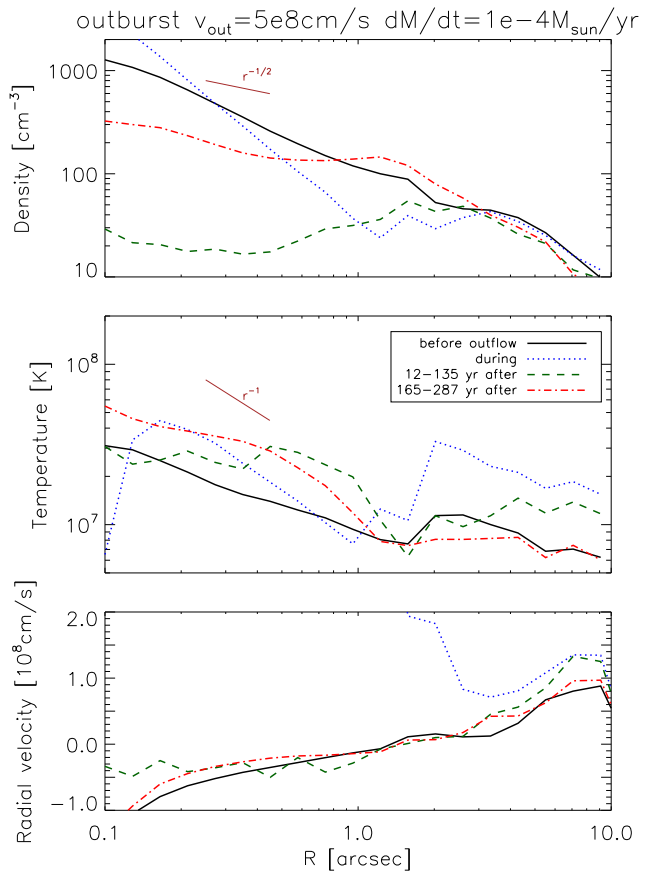}
	\caption{Radial profiles of different gas properties (as in Fig.~\ref{fig:if_profiles}) for the fiducial outburst run at different times -- see inset legend for details.  While the outflow is active the velocity in the inner region takes the value given as input, out of the plot scale.}
	\label{fig:ob_fid_profiles}
\end{figure}

In this subsection we present the results of the simulations in which an outflow lasting 300 yr was added in the model.
We concentrate first on a fiducial model with an outflow rate of $10^{-4}\msun\,$yr$^{-1}$ and a velocity of 5,000 km/s (see \S~\ref{sec:ob}).
Figure~\ref{fig:ob_fid_profiles} shows the radial profiles for this run at different times.  To avoid  fluctuations, each profile is created averaging over 21 snapshots from the simulation, which encompass a period of 122 yr.  Thus they are not to be taken as   instantaneous states of the system, but as the averages over those periods.  We show profiles taken just before the onset of the outflow, during the outflow, and 12--135 and 165--287 yr after the outflow.

Before the outburst we have  an unperturbed state in which the density and temperature profiles increase smoothly towards smaller radii.  The only substructure present is that produced outside the inner arcsecond by the stellar winds.  Notice that these profiles differ from the control run profiles of the previous section (see Fig.~\ref{fig:if_profiles}) -- here we have a steeper density  and a shallower temperature profile.  The difference is due to the different times at which the profiles are taken.  For the instantaneous feedback profiles we took the system at $t \approx 1100\,$yr, which, given the stellar positions along their orbits, corresponds to the present epoch.  For the outburst simulations, that present time is 100 yr after the outburst finishes, so the ``before outflow" state is at $t \simlt 700\,$yr.  The different stellar configurations at different times do produce different gas profiles, as it can be inferred from the capture rate versus time plot, Fig.~\ref{fig:if_accretion}.

During the outburst the density profile is very close to an $R^{-2}$ power-law, as expected for dynamics completely dominated by an outflow.  The temperature shows  the formation of a shock front  where the outflow from the inner region meets the stellar winds.  In the century after the outflow is over,  the density in the inner region is strongly reduced compared to its pre-outflow values, but after another 140 years or so it has mostly recovered.
Unlike the instantaneous feedback mode, when the outburst is active, there is no equilibrium between inflow and outflow.  That results in a positive radial velocity, as seen in the lower panel of the figure.  Either before or after, the inner arcsecond velocity field is dominated by the inflow of the stellar wind material.

In this fiducial run, the density profile gets close to $R^{-1/2}$  during the $165-287\,$yr time range, at which the temperature profile is almost steep enough to fulfil the observational constraints.

\begin{figure}
	\centering \includegraphics[width=0.5\textwidth]{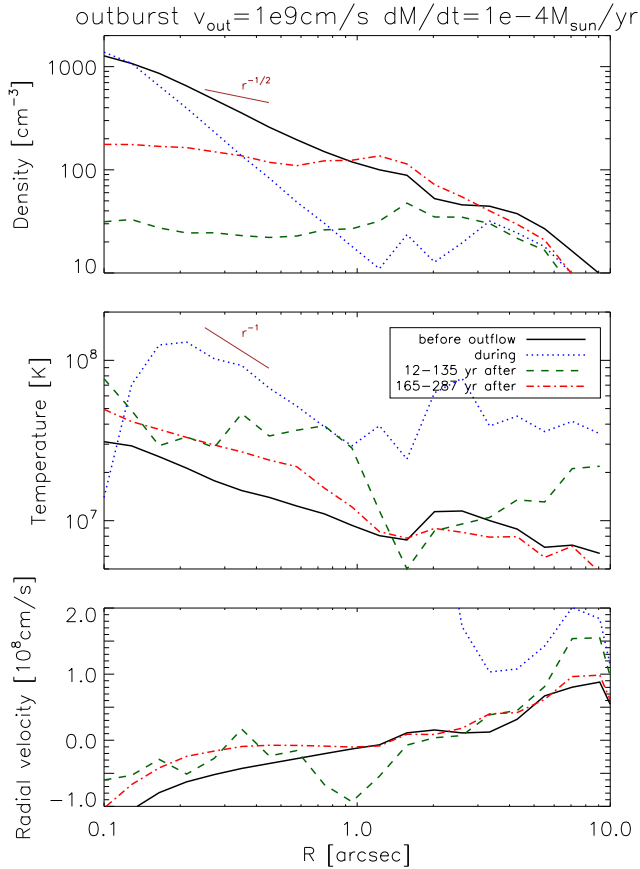}
	\caption{Same as Fig.~\ref{fig:ob_fid_profiles}, but for the run with an outflow velocity $10^9$ cm/s.}
	\label{fig:ob_1e9_profiles}
\end{figure}

A second simulation had the same characteristics, except that the outflow velocity was twice larger, $10,000\,$km/s.  Figure~\ref{fig:ob_1e9_profiles} shows the corresponding profiles.  Qualitatively we see a similar behaviour, but there are some interesting differences.
During the outflow the density is  lower than in the fiducial simulation, which is expected, as $\dot M_{\rm out}$ is the same as before and the outflow velocity is higher.  Notice also that the influence of the outflow reaches farther out into the region dominated by the stellar winds, with higher temperature and velocity, and lower density, for $R\simgt2"$ while the outflow is active.  Moreover, the temperature reached by the gas during and after the outflow are higher, which is consistent with the higher amount of energy injected into the system.  The post-outburst density profile is always too shallow to reach the observational constraints.


\begin{figure}
	\centering \includegraphics[width=0.5\textwidth]{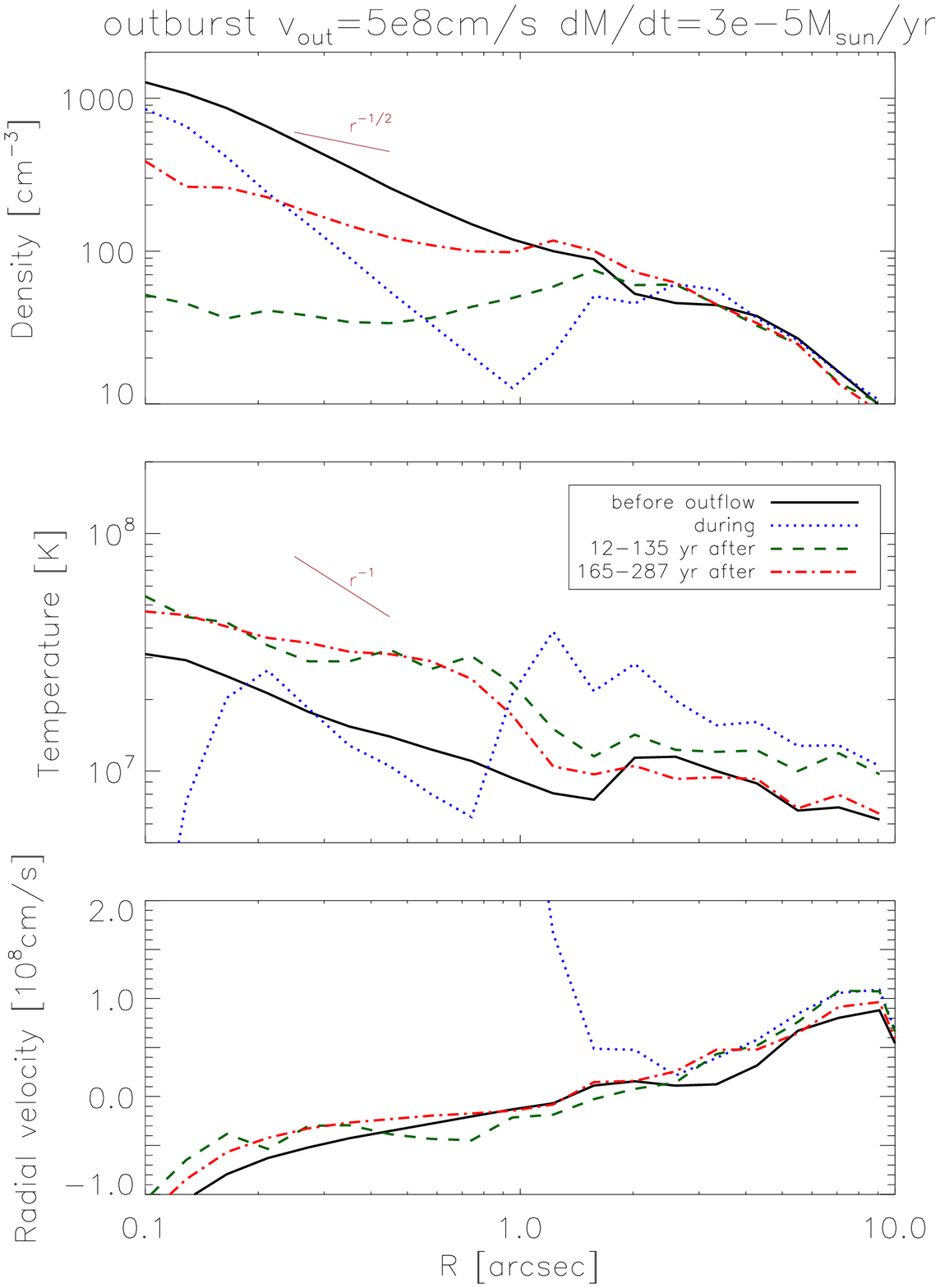}
\caption{Same as Fig.~\ref{fig:ob_fid_profiles}, but for the run with an outflow rate of $3\times10^{-5}\,\msun\,$yr$^{-1}$.}
	\label{fig:ob_3e-5_profiles}
\end{figure}

A third simulation had the same properties of the fiducial run, but three times lower mass-loss rate.  The profiles are shown in Fig.~\ref{fig:ob_3e-5_profiles}.  As expected, the density is lower during the outflow, and its influence does not reach as far out as in the fiducial run.  Moreover, the density profile ``recovers" more quickly in this run.  However, the temperature profile remains too shallow in most of the radial range of interest, so it cannot meet the observational constrains.

\begin{figure}
	\centering \includegraphics[width=0.5\textwidth]{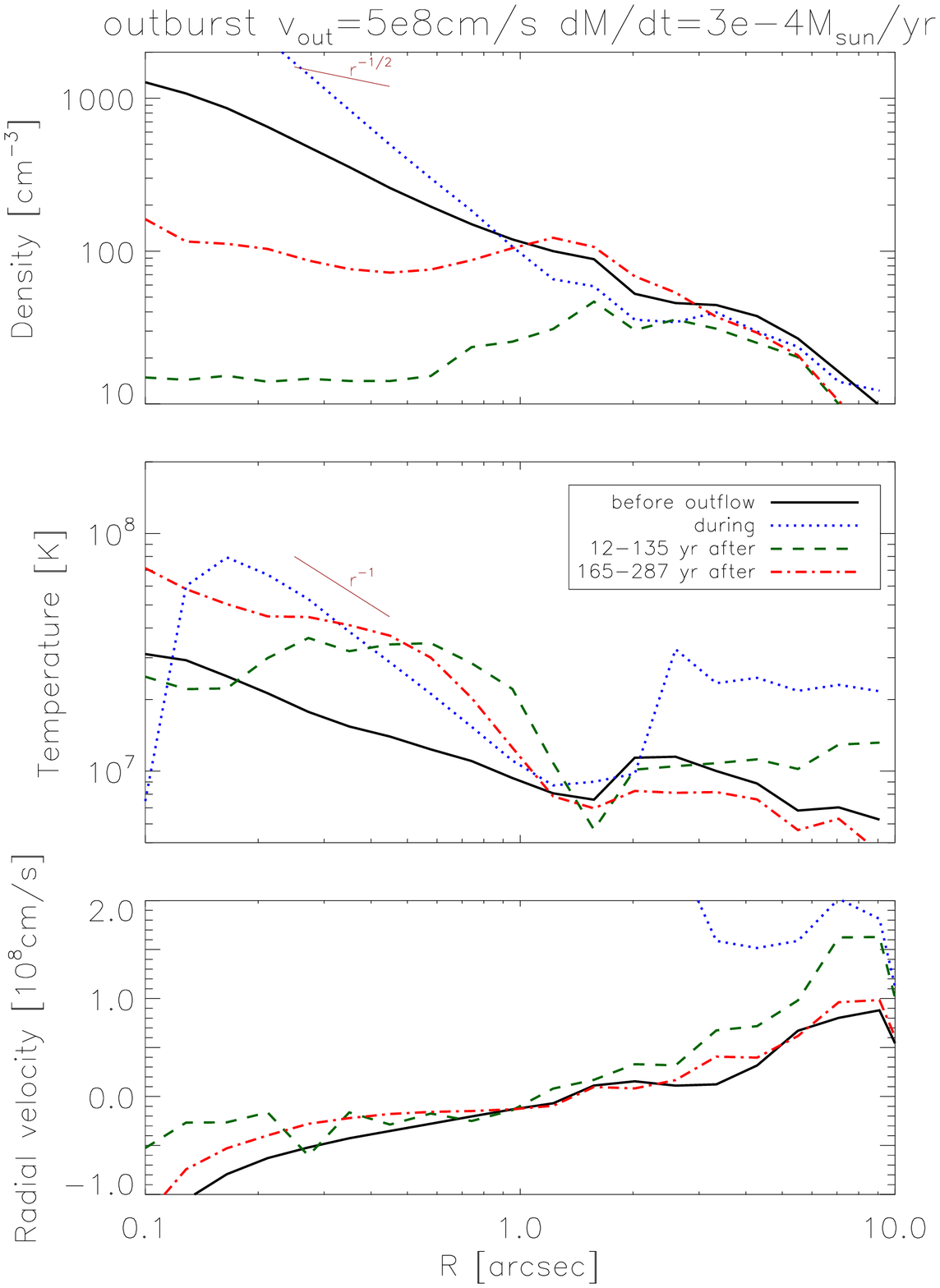}
\caption{Same as Fig.~\ref{fig:ob_fid_profiles}, but for the run with an outflow rate of $3\times10^{-4}\,\msun\,$yr$^{-1}$.}
	\label{fig:ob_3e-4_profiles}
\end{figure}

The fourth simulation increased the mass outflow rate in a factor 3 with respect to the fiducial run.  In this case there is the expected increase in the density and decrease in the temperature during the outflow.  After the outflow is over, however, there are no important differences with the fiducial run, except that in this run the density profile remains  too shallow to reach the observational constraints over the time-frame of interest.

\begin{figure}
	\centering \includegraphics[width=0.5\textwidth]{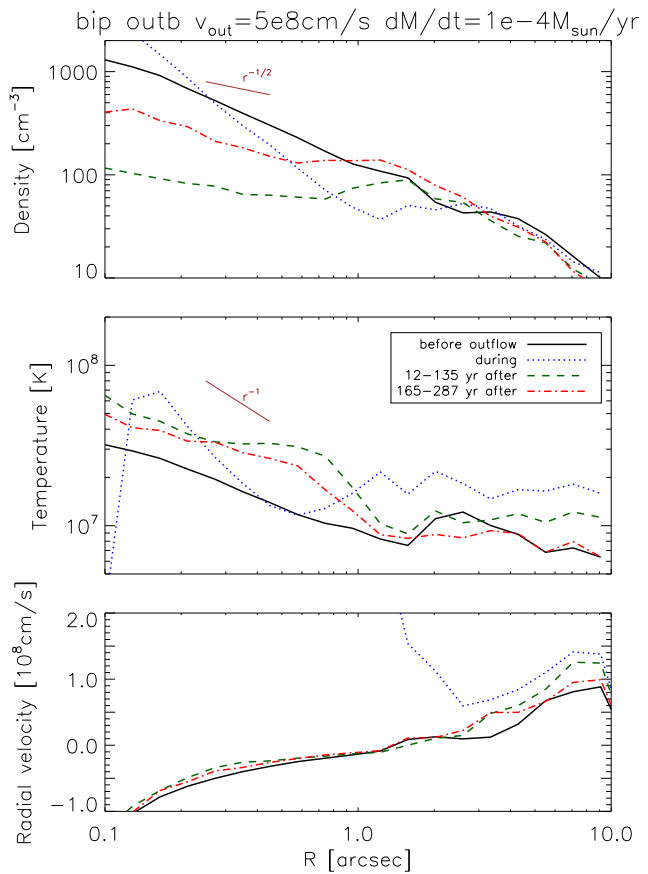}
	\caption{Density (top), temperature (middle), and radial velocity (bottom) profiles for the outburst run with the fiducial values, but a bipolar geometry, at different times -- see inset legend for details.  While the outflow is active the velocity in the inner region takes the value given as input, out of the plot scale.}
	\label{fig:ob_bip_profiles}
\end{figure}

\begin{figure}
	\centering \includegraphics[width=0.5\textwidth]{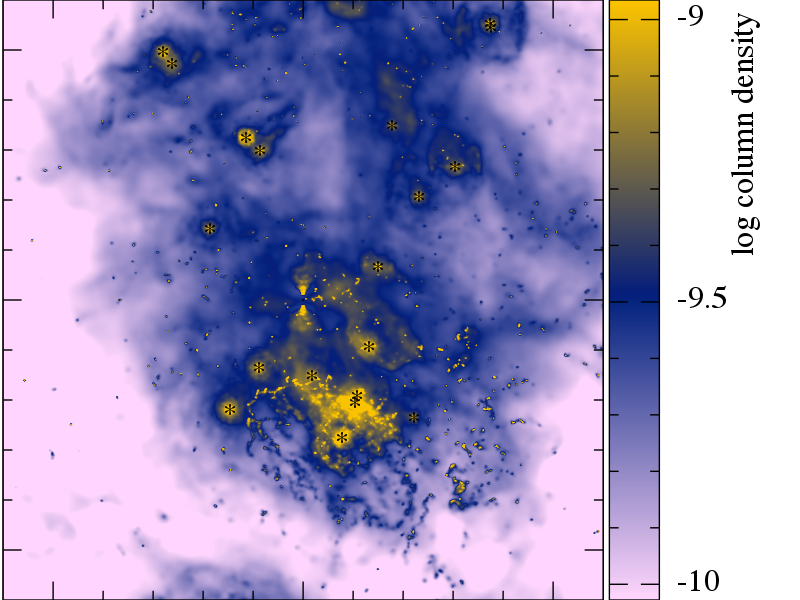}
	\caption{Column density map of the simulation with a bipolar outflow, while the outflow is on.  The map is $12"$ a side and centred on the black hole.  Notice the conical feature at the centre, but the lack of substantial asymmetry further out.}
	\label{fig:ob_biprot_densmap_on}
\end{figure}

While all simulations so far had an isotropic outflow, the fifth simulation shown in Fig.~\ref{fig:ob_bip_profiles} shows a variation of the fiducial run but with the outflow confined to a bipolar cone with half-opening angle of $15\deg$.  Similar as with the instantaneous feedback, this run shows the effect of the outflow to be milder than in the isotropic case.  In this run, the density profile at late times is roughly consistent with the observational constraints, but the temperature profile is too shallow.   Figure~\ref{fig:ob_biprot_densmap_on} shows a density map of this model while the outflow is active.  Only a small-scale ($R\simlt1"$) conical feature is visible, which quickly disappears once the outflow is turned off.

\begin{figure}
	\centering \includegraphics[width=0.5\textwidth]{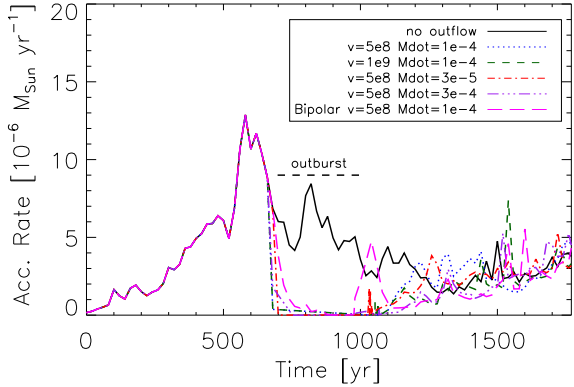}
	\caption{``Capture rates" as a function of time for the different runs using the outburst mode.  The solid, black line shows the control run, without outflow.  The coloured lines show the results of different runs, using different outflow properties.}
	\label{fig:ob_accretion}
\end{figure}

Figure~\ref{fig:ob_accretion} shows the capture rate for the simulations with an outburst presented in this section, together with the control run (without feedback of any kind, same as in Fig.~\ref{fig:if_accretion}).  In most outburst simulations, the capture rate goes down to zero while the outburst is active.  Then there is a period of 200--300 years during which the accretion is still suppressed, to come back to values compatible with the control run.  That recovery time is roughly  the free-fall time of the gas from the location of the nearby stars, $\approx 2"$ away from \sgra.  The only exception to this behaviour is shown by the bipolar run, in which case the capture rate is not completely suppressed and it increases rapidly after the outburst is over.

\subsection{Fraction of hot gas}

  A recent discovery is the high-temperature gas in the so-called `\sgra\ halo' region \citep{Wang13}, defined by   projected radii in the range $2$--$6''$.  This hot gas could be the result of the outflows we are modelling in this study.

\begin{figure}
	\centering \includegraphics[width=0.5\textwidth]{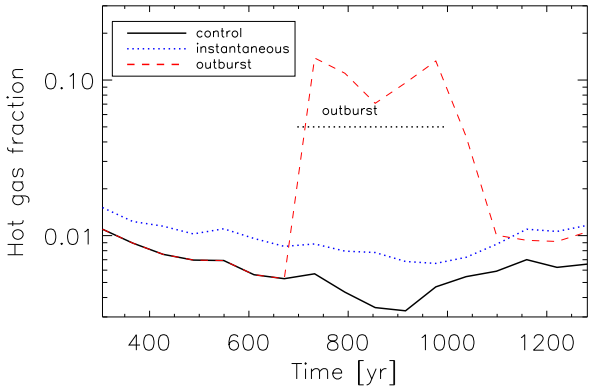}
	\caption{Mass fraction of hot gas as a function of time for different runs. We show the control run, and the simulations that produce the most hot gas for both the `instantaneous feedback' and `outburst' outflow modes. }
	\label{fig:hot_fraction}
\end{figure}

 We took the different simulations and calculated the mass fraction of hot ($T>10^8\,$K) gas in that region.  The results are plotted in Fig.~\ref{fig:hot_fraction} as a function of time for a few selected simulations.  The control run, without any outflows, shows a fraction of hot gas that varies within 0.3--1\%.  This variation, as that of the accretion rate (Fig.~\ref{fig:if_accretion}), arises from the variations in the stellar configuration due to their orbital motion.  The models with an `instantaneous feedback' show a similar pattern, but a slightly higher hot gas fraction -- the dotted line in the figure shows the largest effect we achieved in our runs, with $v_8 = 10$.  In the simulations that include an `outburst' for the feedback we see that the fraction increases substantially during the outflow (time interval marked by a horizontal line), reaching values of around 10\% for the simulation with outflow velocity $v_8 = 10$ and mass loss rate $10^{-4}\msun\,$yr$^{-1}$.    This effect, however, quickly disappears, the mass fraction of hot gas goes back to its pre-outflow value in a hundred year period after the outflow is over.  This is expected, as hot gas has a very large sound speed and quickly leaves this region.

\begin{figure*}
	 \includegraphics[width=0.49\textwidth]{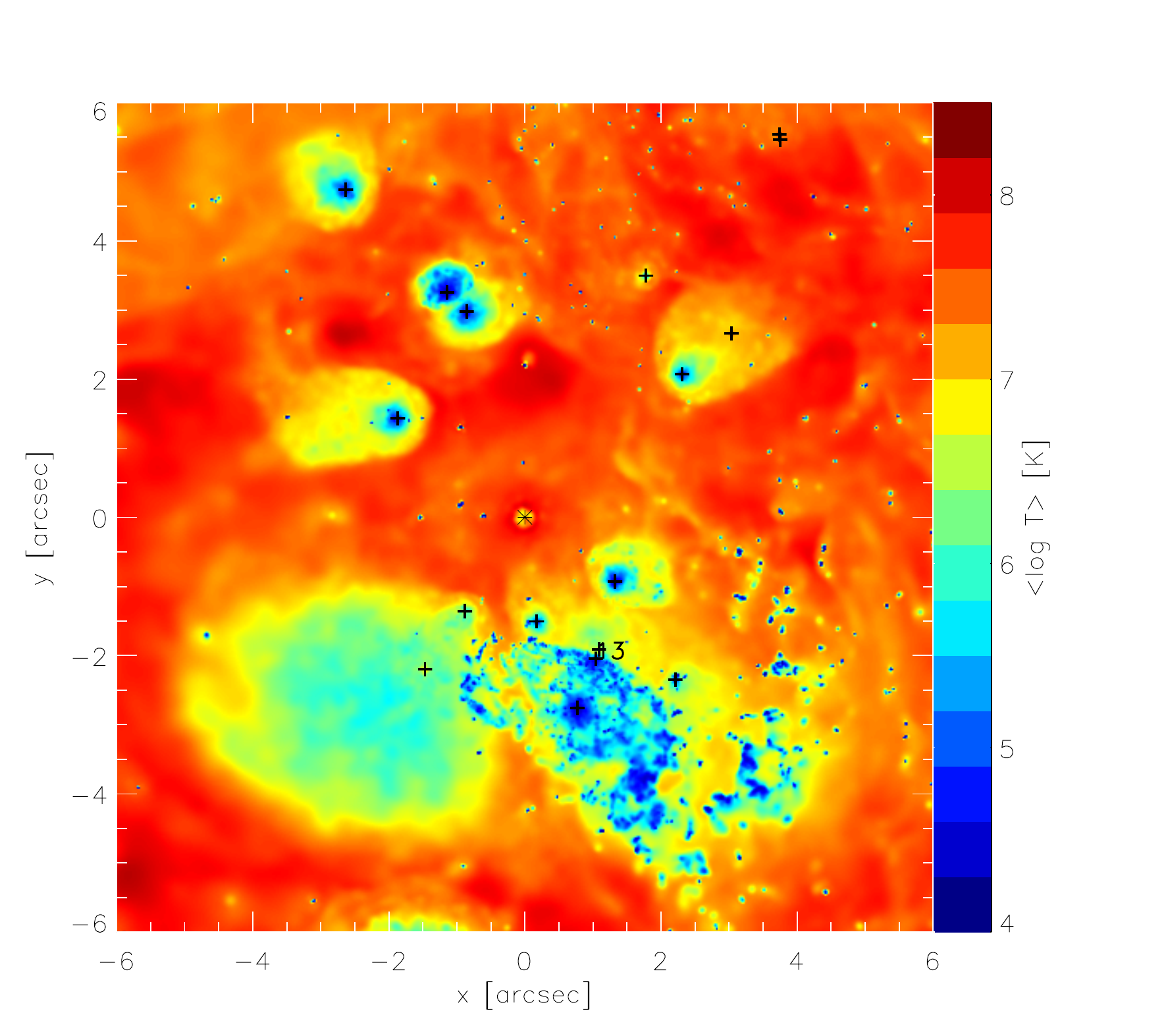}
	 \includegraphics[width=0.49\textwidth]{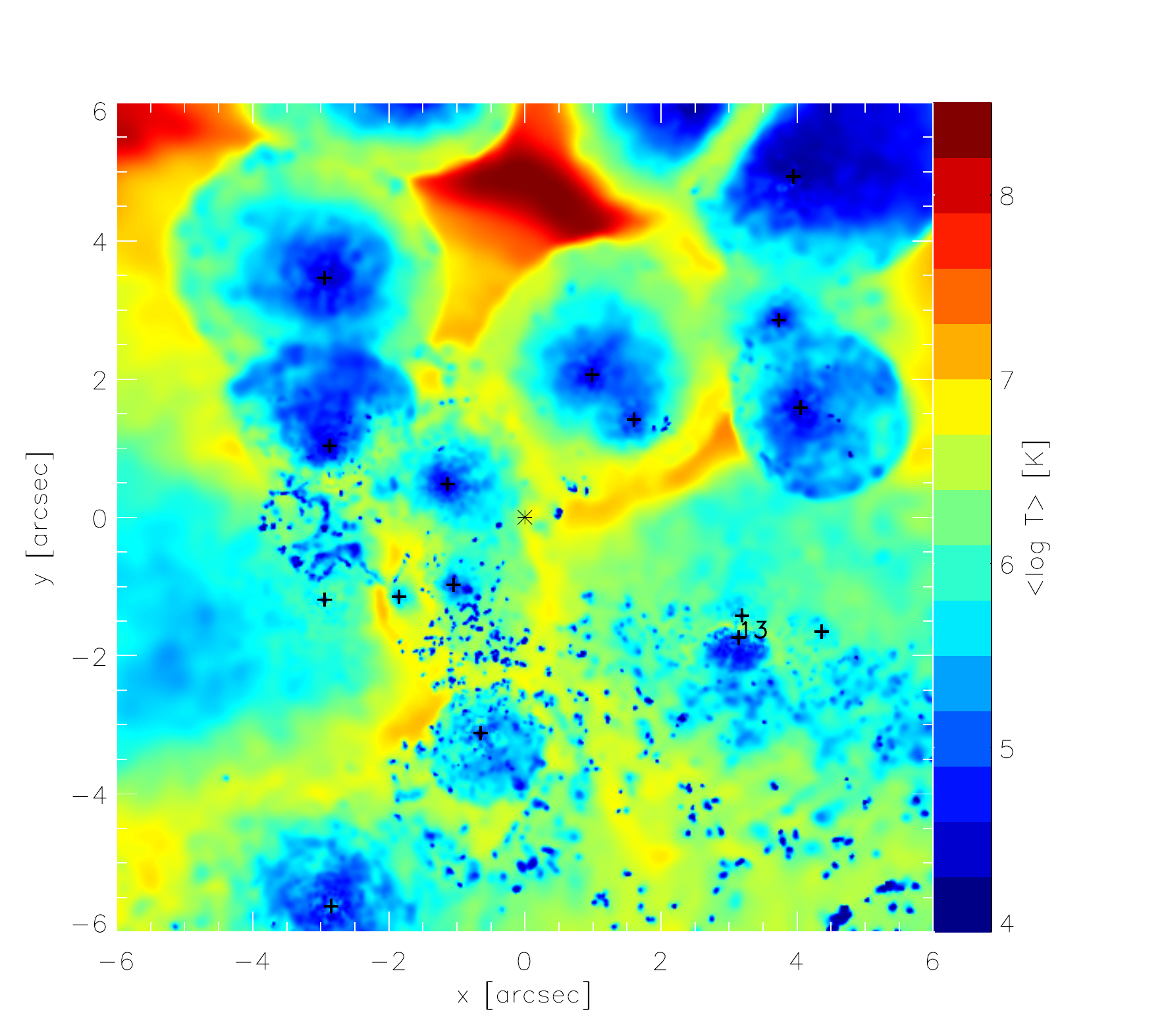}
	\caption{Temperature maps of the simulation with an outburst of  $v_8 = 10$ and mass loss rate $10^{-4}\msun\,$yr$^{-1}$.  The left panel shows a time when the outflow is active, and a noticeable amount of hot gas is distributed throughout the computational domain.  The right panel shows the current time, after the the outburst has finished and only a small fraction of hot gas remains.  The maps are $12"$ a side and centred on the black hole.}
	\label{fig:burst_1e-4_1e9_132_tempmap}
\end{figure*}

Figure~\ref{fig:burst_1e-4_1e9_132_tempmap} shows the temperature maps of the simulation with the strongest outflow, both while it is active and at the present time, after the outburst has stopped.  It is clear from the left map that the hot gas is distributed throughout the inner computational domain while the outflow is active, and it is only locally hindered by interactions with {\it stellar} outflows.  This gas, as most of the gas outside a 1--$2"$ radius in all of our models,  is outflowing from the computational domain.
As discussed above, once the outburst is over, the hot gas quickly leaves the system, and most of the remaining gas is cool, with temperatures comparable to those in the control run (cf. Fig.~\ref{fig:control_tempmap}, right).

\section{Discussion}

In this study, we have improved our previous models of the Galactic centre gas dynamics \citep{CuadraEtal08}.  While those models roughly reproduced the capture rate at the Bondi radius, they were not completely physical in that they ignored the expected presence of an outflow from the inner accretion flow.   

We modelled two different types of outflow:  (i) an instantaneous feedback, in which material is constantly ejected as it approaches the black hole, and (ii) an outflow which is stronger, but active for a limited time. This latter situation is applicable to sources that went through a recent enhanced black hole accretion episode, which may be the case for \sgra. In both cases we found important changes in the dynamics near the capture radius and in the accretion rate.  We concentrated mostly on isotropic black hole outflows, but also tested outflows with moderate beaming.  In those cases we found the effect of the outflow was reduced, but still noticeable in our models.  While the models are specifically tailored for the Galactic centre, our conclusions  are also relevant for other  LLAGN,  as explained below.

The instantaneous feedback mode shows that the gas at the Bondi radius does not uniquely set the rate at which gas is captured.  Models with different outflow velocities (within the relatively narrow range we explored) give accretion rates that differ by a factor 5, despite nearly identical gas properties at $R\simgt1"$.  
On the other hand, the outburst mode shows that our current understanding of \sgra's accretion flow, based on steady-state models, should be revised.  If, as believed, \sgra\ was much more active a couple of centuries ago, an outflow of similar power to the inferred luminosity ($\sim10^{39}\,$erg/s) could still be affecting the gas we observe at $R \simgt 0.1"$.  The accretion rate and gas density at the Bondi radius could still be increasing to pre-outburst values, while the temperature is decreasing.  These ``complications" add to the ones identified earlier \citep{CuadraEtal08} of variability caused by the orbital motion of the stars around \sgra\ and by the stochastic clump formation.  

Overall, the Bondi model, which is symmetrical and time-independent, is not able to account for the complexity of the accretion--outflow process around \sgra.  This conclusion is similar to that reached by other authors for  LLAGN.\footnote{It is interesting to note that, while the Bondi model overestimates the accretion rate for the case of \sgra\ and LLAGN, in  galaxy clusters the Bondi estimate gives too low an accretion rate compared to what is measured for its central AGN \citep[e.g.,][]{Cavagnolo11}, but that is a very different regime to the one we study here.}
  \cite{Hillel13} and \cite{Shcherbakov14}  have focused on models for  NGC 3115, which is a nearby ($\approx 10\,$Mpc) LLAGN with a $\sim10^9\msun$ central black hole, meaning that its Bondi radius is easily resolvable \citep[e.g.,][]{Wong14}.  \cite{Hillel13} argue that the Bondi model is not applicable, as there is central pressure produced by the old stellar population (slow) winds, which originate within the Bondi radius. The 1D numerical steady-state model of \cite{Shcherbakov14} also includes the effect of stellar winds and supernovae inside the Bondi radius, plus thermal conduction to transport the energy out.  Both these models conclude that a time-independent solution is not physical, and that ouflows are produced.  Notice that the physical situation is somewhat different in NGC 3115 compared to our Galactic centre, as in the latter the stellar wind sources are located roughly {\it at} the Bondi radius, and there are no significant stellar sources within \citep[but see][]{Loeb04}.  Nevertheless, the role of the stellar winds in NGC 3115 is comparable to the outflow included in our model. 

Due to numerical reasons, our models did not include thermal conduction.  The effect of thermal conduction would be to flatten somewhat the density and temperature profiles \cite{Shcherbakov14}, and it would be mostly circumscribed to the region within the inner arcsecond.  It appears that such a flattening would help to reconcile the post-outburst profiles with the observational estimates, so that is in principle an interesting avenue for future research.  However, the strong magnetic field likely present in the Galactic centre \citep{Eatough13} would suppress the conduction.

We compared our models with the observational constraints obtained from the recent 3\,Ms {\it Chandra} observation of \sgra \citep{Wang13}.  None of our instantaneous feedback models fits the data, but we find that an isotropic outflow, active $\sim 200\,$yr ago with a rate of $10^{-4}\msun\,$yr$^{-1}$ and a velocity of 5,000 km/s, roughly reproduces the currently observed density and temperature profiles.
 We also rule out any such event happening in the last $\sim100\,$yr.    
Our method has therefore the potential to constrain the past activity of \sgra, complementing the X-ray echo method which suffers from ambiguity due to the unknown 3D location of the reflecting molecular clouds  \citep[e.g.,][]{Clavel13}. 
  These comparisons should be taken at a qualitative level only, as they are based on power-law fits to the observed data and on radial profiles from the simulation data.  In a forthcoming paper, we will use the resolved data from both sources to perform a more robust comparison.

Our models show that some gas with very high temperatures remains in the vicinity of \sgra\ for a period of time after the outflow is over.  While that cannot explain at face value the recent Chandra observations of such hot gas \citep{Wang13}, it is worth noticing that ours is probably a lower limit.  Due to numerical limitations, in our simulations there is an outer boundary condition of free flow at $12"$, while in reality higher density gas on the Circum-nuclear disc outside that boundary is likely to help in keeping that gas confined.  Moreover, we have explored rather mild outflows -- for stronger outbursts the amount of hot gas and its temperature would be higher and likely to linger on for longer.  Further studies of such outflows and comparisons with observational data will help us constrain the past activity of \sgra.

\section*{Acknowledgements}

Density maps were created with {\sc splash}, by \cite{Price07}.
We thank the anonymous referee for useful comments that greatly helped us to improve the presentation of the paper.
JC acknowledges support from CONICYT-Chile through FONDECYT (1141175),
Basal (PFB0609) and Anillo (ACT1101) grants, and the warm hospitality at Leicester.
Theoretical astrophysics research in Leicester is supported by an STFC Rolling
Grant.   QDW acknowledges the support by NASA via the SAO/CXC grant TM3-14006X.
This work was performed using the Geryon computers at the
Centre for Astro-Engineering UC, funded by Basal PFB-06,  QUIMAL 130008, and Fondequip AIC-57.

\bibliographystyle{mnras}
\bibliography{nayakshin}

\label{lastpage}

\end{document}